\documentclass[12pt]{iopart}

\usepackage{graphicx}
\usepackage{tabularx}
\usepackage{hyperref}
\usepackage{amsmath}
\usepackage{mathrsfs}
\usepackage{amsfonts}

\begin{document}

\title{Small electron polarons bound to interstitial tantalum defects in lithium tantalate}

\author{Anton Pfannstiel~$^{1}$, Tobias Hehemann~$^{1}$, Nils A. Schäfer~$^{2}$, Simone Sanna~$^{2}$, Yuriy Suhak~$^{3}$, Laura Vittadello~$^{1, 4}$, Felix Sauerwein~$^{1}$, Niklas D\"omer~$^{1}$, Julian Koelmann~$^{1}$, Holger Fritze~$^{3}$ and Mirco Imlau~$^{1, 4, *}$}

\address{$^{1}$ Institute of Mathematics/Informatics/Physics, University of Osnabr\"uck, Barbarastra\ss{}e 7, D-49076 Osnabr\"uck, Germany}
\address{$^{2}$ Institut für Theoretische Physik and Center for Materials Research (ZfM/LaMa), Justus-Liebig-Universit\"at Gie\ss en, Heinrich-Buff-Ring 16, D-35392 Gie\ss en, Germany}
\address{$^{3}$ Institut für Energieforschung und Physikalische Technologien, Technische Universit\"at Clausthal, Am Stollen 19 B, D-38640 Goslar, Germany}
\address{$^{4}$ Center for Cellular Nanoanalytics, , University of Osnabr\"uck, Barbarastra\ss{}e 11, D-49076 Osnabr\"uck, Germany}
\ead{mirco.imlau@uni-osnabrueck.de}
\vspace{10pt}
\begin{indented}
\item[]January 2024
\end{indented}

\begin{abstract}
The absorption features of optically generated, short-lived small bound electron polarons are inspected in congruent lithium tantalate, LiTaO$_3$ (LT), in order to address the question whether it is possible to localize electrons at interstitial Ta$_{\rm V}$:V$_{\rm Li}$ defect pairs by strong, short-range electron-phonon coupling. Solid-state photoabsorption spectroscopy under light exposure and density functional theory are used 
for an experimental and theoretical access 
to the spectral features of small bound polaron states and to calculate the binding energies of the small bound Ta$_{\rm Li}^{4+}$ (antisite) and Ta$_{\rm V}^{4+}$:V$_{\rm Li}$ (interstitial site) electron polarons. As a result, two energetically well separated ($\Delta E \approx 0.5$ eV) absorption features with a distinct dependence on the probe light polarization and peaking at 1.6 eV and 2.1 eV are discovered.
We contrast our results to the interpretation of a single small bound Ta$_{\rm Li}^{4+}$ electron state with strong anisotropy of the lattice distortion and discuss the optical generation of interstitial Ta$_{\rm V}^{4+}$:V$_{\rm Li}$ small polarons in the framework of optical gating of Ta$_{\rm V}^{4+}$:Ta$_{\rm Ta}^{4+}$ bipolarons. 
We can conclude that the appearance of carrier localization at Ta$_{\rm V}$:V$_{\rm Li}$ must be considered as additional intermediate state for the 3D hopping transport mechanisms at room temperature in addition to Ta$_{\rm Li}$, as well, and, thus, impacts a variety of optical, photoelectrical and electrical applications of LT in nonlinear photonics. Furthermore, it is envisaged that LT represents a promising model system for the further examination of the small-polaron based photogalvanic effect in polar oxides with the unique feature of two, energetically well separated small polaron states.
\end{abstract}

\section{Introduction}\label{sec:introduction}
Small electron polarons, i.e. electrons self-trapped at essentially one site in condensed matter by strong, short-range interaction of the carrier with the surrounding lattice~\cite{Emin2013}, determine the electrical and optical properties of polar oxides, such as lithium tantalate, LiTaO$_3$ (LT), to a large extend. The small polaron binding energy is considerably enhanced by the extra potential of lattice defects such as Ta$_{\rm Li}$ antisites in non-stoichiometric LT, resulting in a complete lattice site localization at even modest irregularities, i.e. yielding small \textit{bound} polarons. Only thermally activated hopping then allows polaron transport to neighbouring sites or, alternatively, the optical absorption of photons. The latter gives rise to large absorption cross sections (in the order of $10^{-22}$ m$^2$) of the bound states that range over a broad spectral range ($\sim 1.0$ eV) and can be used to probe the defect structure of LT by optical means. 

Lithium tantalate grown via the Czochralski method~\cite{Ballmann1967, Fedulov1965, Uecker2014} from a melt of 49.0 mol\% LiO$_2$~\cite{Barns1970, Fedulov1965} (Ta$_2$O$_{\rm 5}$/(Ta$_2$O$_{\rm 5}$+Li$_2$O) =0.4875~\cite{Miyazawa1971} and [Li]/([Li]+[Ta])=0.485~\cite{He2008}),
commonly called 'congruent LiTaO$_3$' (c-LT), features an intrinsic defect structure that can be described using the Li-vacancy model~\cite{Barns1970} (widely accepted for the isomorphic and isoelectric ferroelectric polar oxide lithium niobate, LiNbO$_3$ (LN)~\cite{Iyi1992, Bluemel1994}), although a pronounced sample dependence has been found in the past~\cite{Vyalikh2018}.
In accordance with investigations from first principles, the presence of both lithium vacancies V$_{\rm Li}$~\cite{He2016} and Ta$_{\rm Li}$~\cite{Vyalikh2018} antisite defect clusters can be considered to ensure charge compensation. The ability of the LT structure to accommodate an excess of Ta is a result of the strong covalency of the Ta-O bond~\cite{Peterson1968}. As point of reference, and only assuming Ta$_{\rm Li}$ antisites in the defect structure, an antisite concentration of $c_{\rm antisite}\approx 10^{26}$ m$^{-3}$ is estimated for c-LT \cite{Merschjann2008}. The formation of small bound polarons upon thermal and/or optical excitation of free carriers can thus be expected and was studied in direct comparison to the features of LN by means of first-principles calculations and ultrafast transient absorption (pump-probe) and femtosecond upconversion luminescence spectroscopy~\cite{Krampf2021}, only recently. Experimental evidence was deduced for the presence of small free Ta$_{\rm Ta}^{4+}$ electron polarons, small bound Ta$_{\rm Li}^{4+}$ electron polarons, bound O$^-$-hole polarons, and Ta$_{\rm Ta}^{4+}$:Ta$_{\rm Li}^{4+}$ bipolarons. Also the coupling of Ta$_{\rm Ta}^{4+}$ electron and O$^-$-hole polarons via Coulomb interaction yielding self-trapped excitons~\cite{Krampf2021, Song1996} could be concluded, thus resembling the small electron and hole polaron features of LN in a general manner~\cite{Schirmer2009,Schirmer2006}. 

The defect model for LT  further allows the presence of another defect center, the \textit{interstitial} Ta$_{\rm V}$ defect, 
i.e., the incorporation of Ta on a cation site that is usually kept unoccupied in the case of the stoichiometric composition~\cite{Barns1970}. 
The Ta$_{\rm V}$ defect can therefore be regarded as alternative option to reach a Ta excess in non-stoichiometric LT. For charge compensation, it will be accompanied by five Li vacancies V$_{\rm Li}$, one of which on the direct neighboring Li lattice site. Thus, a Ta$_{\rm V}$:V$_{\rm Li}$ defect pair directed along the polar ${\bf c}$-axis is always present and the sequence of metal atoms along the polar $\bf c$-axis is changed from ... Ta $|$ $\square$ $|$ Li $|$ Ta $|$  $\square$ $|$ Li ... ~\cite{Abrahams1967} to ... Ta $|$ $\square$ $|$ Li $|$ Ta $|$ Ta$_{\rm V}$ $|$ V$_{\rm Li}$ ... (here, $|$ denote the oxygen planes and $\square$ unoccupied sites).  
Structural investigations have shown, that it is reasonable to assume a mixture of Ta$_{\rm Li}$ antisite defects and interstitial Ta$_{\rm V}$:V$_{\rm Li}$ defect pairs compensated by V$_{\rm Li}$ in c-LT with a tendency of a larger antisite concentration~\cite{Barns1970}. 
 However, there is still a lack of experimental evidence for Ta$_{\rm V}$ sites in LT, and particularly for the formation of small bound Ta$_{\rm V}^{4+}$:V$_{\rm Li}$
electron polarons. Furthermore, calculations of respective formation and binding energies as well as of optical spectra are missing in literature, so far. Thus, the knowledge about the microscopic origin of broad band optical absorption in the near-infrared spectral region, commonly assigned to the small bound Ta$_{\rm Li}^{4+}$ electron polaron, is strongly limited. Without this knowledge, it is not known, whether interstitial defect centers exist at all in LT and how they may contribute via electron capture to the optical absorption. Furthermore, there is no knowledge about the possible interplay between small polarons bound at different intrinsic lattice defect sites, about the question whether both types of bound states can exist simultaneously and to which extent they contribute to the charge transport mechanisms in LT. 

We have addressed these questions, both, experimentally and theoretically by polarized absorption spectroscopy of c-LT single crystals, and first principles calculations, respectively. Small bound electron polarons were prepared as long-lived states by the combination of light exposure used for the gating of Ta$_{\rm V}^{4+}$:Ta$_{\rm Ta}^{4+}$ bipolarons, and by cooling the c-LT single crystal to temperatures below T=100 K in order to slow-down hopping transport and a subsequent decay via recombination with Ta$_{\rm Ta}^{4+}$. We find two spectrally well separated absorption bands using different probe light polarizations with peak energies of the absorption maxima that coincide well with the theoretical expectations. The relative strengths may be attributed to a larger number density of Ta$_{\rm Li}^{4+}$ assuming similar small polaron absorption cross sections for bound polarons at antisites and interstitial defects. It is possible to conclude that antisite and interstitial polarons coexist as separate and mutually independent quasiparticles in c-LT, but also that migration of antisite tantalum atoms to adjacent, unoccupied oxygen octahedra is at the origin of the two discovered absorption finger prints. The impact of our findings for the charge transport, in particular for the strength and temporal dynamics of the small-polaron based bulk photogalvanic effect, is discussed.

\section{Methodology}\label{sec:methodology}
\subsection{Congruent lithium tantalate crystals}
Lithium tantalate, LiTaO$_3$, single crystal plates were prepared from a 3-inch $x$-cut wafer (Precision Micro-Optics Inc., USA) of a boule that was grown by Czochralski method from the congruent melt composition. The entrance faces of the samples with dimensions of $6\times 5$ mm$^2$ (polar $\bf c$-axis parallel to the 5 mm side) and thickness of $d=500 \mu$m were polished to optical quality, so that losses due to light scattering of an incident laser beam was not observable. The samples are subjected to heat treatment at 1\,200 $^{\circ}$C for a duration of ten hours in a tube furnace (Carbolite Gero GmbH, Germany) under reducing ${\rm Ar}+5\%{\rm H}_2$ atmosphere and a heating and cooling rate of 1 K/min (for details see Ref.~\cite{Schulz2013}). Thus, the conditions for reduction annealing were chosen according to Ref.~\cite{Alexandrovski1999} yielding 'slight' (and not 'heavily'~\cite{Kappers1985}) reduced c-LT samples.

\subsection{Polarized absorption and infrared spectroscopy }
A commercial two-beam photospectrometer (UV-3600, Shimadzu Deutschland GmbH) equipped with a combination of deuterium $\&$ halogen lamp and double grating monochromator is used for polarized absorption spectroscopy of the c-LT samples in the ultraviolet/visible/near-infrared (UV/VIS/NIR) spectral range of $250-2\,500$ nm (spectral resolution $<5$ nm FWHM at 550 nm and 800 nm, probe light detection with a multialkali photomultiplier in the UV/VIS and an InGaAs detector in the NIR).
A closed cycle Helium cryostat (RDK 10-320, Leybold GmbH) equipped with optical quartz-windows and placed in the sample compartment of the spectrometer served for cooling the crystal samples to temperatures of about $T=80$ K. The sample plates were loosely fixed to the window-like opening of the cooling finger (copper) with heat-conducting paste, thus enabling transmission measurements under unclamped conditions. Illumination of the sample ($\lambda=488$ nm, area of exposure $\approx 1 $mm$^2$, intensity up to $I=6\cdot10^{5} $W/m$^2$) was realized with the unexpanded beam of a continuous-wave optically-pumped semiconductor (OPSL) laser (Genesis CX488-2000, Coherent Corp.) that was adjusted by a silver mirror nearly co-linear to the probe beam. The detector's chamber was closed during laser exposure to avoid damage to the detectors and the reference beam of the spectrometer remained empty throughout our studies.
A Glan-Thompson polarizer (B. Halle Nachfl. GmbH) in front of the sample chamber served for polarization dependent spectroscopic studies: ${\bf E}\parallel{\bf c}$- and ${\bf E}\perp{\bf c}$-axis. 
Infrared spectroscopy was additionally applied in order to study the characteristics of the OH$^-$ stretching vibration in the vicinity of a wavenumber of about 3\,480 cm$^{-1}$. Here, a commercial Fourier-transform infrared (FTIR) spectrometer (Bruker Corporation, Vertex 70) was used (unpolarized, without cooling, without laser exposure).

\subsection{Density functional theory}

Atomistic calculations are performed within spin-polarized density functional theory as implemented in the QuantumEspresso computational package \cite{Giannozzi2009}. According to a well-established approach \cite{Friedrich2017, Krampf2021}, we employ norm conserving potentials within the PBEsol parametrization of the exchange and correlation functional \cite{Perdew2008}. A number of valence electrons corresponding to 3, 13 and 6 is considered for Li, Ta and O, respectively. A plane wave basis containing waves up to a kinetic energy of 100 Ry is employed to expand the electronic wave functions. The correct treatment of the $5d$ orbitals of Ta is challenging due to the strong electronic correlation. The strong on-site Coulomb repulsion of such
electrons is accounted for in this work by the LDA$+U$ approach \cite{Anisimov1997} and self-consistently determined $+U$ values. The atomic geometries and total energies are calculated on 3$\times$3$\times$3 repetitions of the rhombohedral unit cells, i.e., with 270 atoms per supercell. A 2$\times$2$\times$2 Monkhorst-Pack k-point mesh \cite{Monkhorst1976} is thereby employed. 
The convergence threshold for forces during structural optimization set to 10$^{-4}$ Ry.
Due to the computational demand, the linear optical response of the defect structures is calculated on smaller 2$\times$2$\times$2 repetitions of the primitive unit cell, i.e., on 80 atoms supercells. Thereby the electronic structure calculated within QuantumEspresso is processed in a perturbative approach by the Yambo package \cite{Sangalli2019} for the calculation of linear response quantities including electron-hole interaction as well as quasi-particle corrections based on the GW formalism. For the latter, the first 1\,000 electronic bands are considered, corresponding to 276 occupied states and 724 empty states. For the solution of the Bethe-Salpeter equation a reduced number of 59 occupied and 54 empty states is considered. This guarantees dielectric function converged till about 5 eV.

\section{Results}\label{sec:results}

\subsection{Ground state absorption}\label{sec:annealed}

Figure~\ref{fig:figure1}a shows the absorption spectra of the as-grown (blue data points) and of the reduction annealed (green data points) congruent lithium tantalate sample in the UV/VIS spectral range from 1.75 eV-4.75 eV measured without polarizer in the optical beam path as an example. 
The as-grown sample reveals a spectrally broad VIS transmission with noticeable increase of absorption from 1.75 eV to 4.3 eV, thus covering the long-wavelength absorption tail commonly described by the Urbach rule~\cite{Kurik1971, Cabuk1999}. The band-to-band absorption from filled oxygen 2$p$ bands to empty Ta 4$d$ bands and exciton formation appears at a photon energy of $\approx 4.5$ eV ($\equiv 275$ nm) in full agreement with the absorption features of nominally undoped c-LT reported in literature~\cite{Alexandrovski1999}. Polarization dependent measurements of the absorption with probe polarization ${\bf E}\parallel {\bf c}$- and ${\bf E}\perp {\bf c}$-axis showed no change with respect to the results of figure~\ref{fig:figure1}. We note that we also found no evidence for characteristic absorption features that may point to extrinsic dopants in any of the absorption spectra (e.g. transition metal impurities reported in Ref.~\cite{Kappers1985} at 4.1 eV and/or Fe$^{2+/3+}$ at 3.0 eV ($\equiv 410$ nm)~\cite{Kraetzig1978}). In contrast, the absorption spectrum is considerably affected by reduction annealing and a characteristic shoulder appears that peaks at about 3.5 eV ($\equiv 350$ nm). Here, the absorption is increased from 0.4 cm$^{-1}$ (as-grown) to around 2.8 cm$^{-1}$ and, thus, by a factor of $\times 3 - \times 5$ smaller in comparison with the absorption changes induced by reduction treatment in 100\% Argon atmosphere~\cite{Kappers1985}. 
In addition, the transmission break-down that is assigned to the optical band gap shows a slight red-shift. Using the Tauc-method and assuming an indirect optical transition, the band gap energies were determined from both transmission spectra revealing E$_{\rm gap}=4.63$ eV for the as-grown and E$_{\rm gap}=4.53$ eV for the reduction annealed c-LT sample. 
Furthermore, the OH$^{-}$ stretching band was measured for the as-grown (blue data points) and the reduction annealed (green data points) and is shown in the inset of figure~\ref{fig:figure1}a. The results are in agreement with literature data, showing a broad absorption band centered at 3485 cm$^{-1}$ with a full width at half maximum (FWHM) of $\approx 35$ cm$^{-1}$ for the as-grown sample. Its shape is slightly asymmetric with a shoulder at 3\,460 cm$^{-1}$ as reported in Refs.~\cite{Koehler2016, Vyalikh2018}. A qualitative similar spectrum with decreased OH$^-$ absorption amplitude is found in the reduction annealed sample in accordance with \cite{Vyalikh2018} for c-LT, but also for lithium niobate \cite{Sugak2010}.

\subsection{Illuminated state}\label{sec:pumped}

\begin{figure}
\centering
\includegraphics[width=15cm]{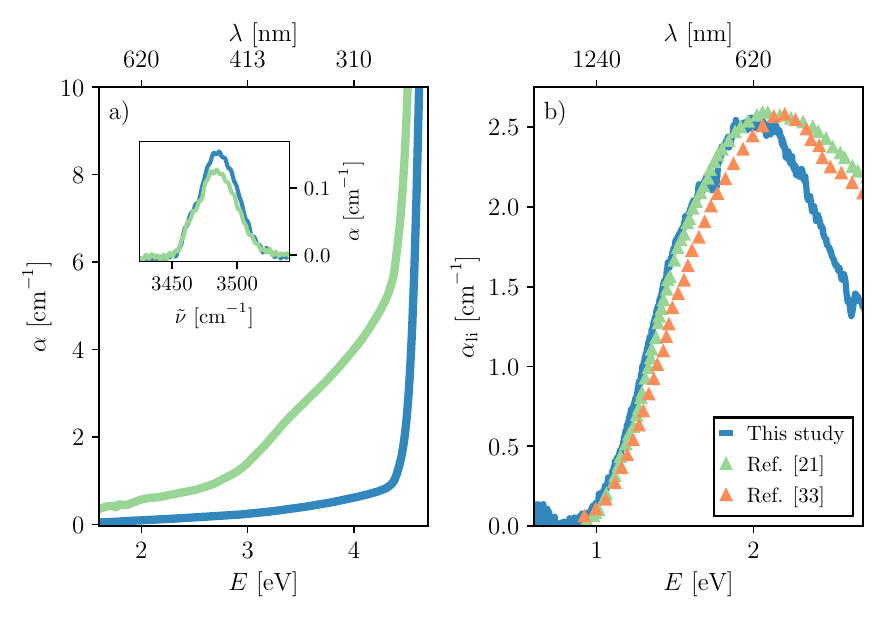}
\caption{a) Absorption spectrum of an as-grown (blue data points) and a reduction annealed (green data points) $x$-cut, congruent lithium tantalate crystal in the visible spectrum, measured with unpolarized probing light and without reflection correction at $T=300$ K. The inset shows the corresponding OH$^{-}$ absorption band in the IR. b) Laser-induced change of the absorption ($Q=I\cdot t\approx 3\cdot 10^{8}$ Ws/m$^2$, $\lambda=488$ nm,  2.54 eV, $T=80$ K) of the reduction annealed c-LT sample of this work (blue), of \textit{acoustic grade} LT published by Kappers \textit{et al.}, 1985~\cite{Kappers1985} (green, amplitude scaled by a factor of 0.08) and of near-stoichiometric (Li ratio: $49.65\%$) LT published by Liu \textit{et al.}, 2004~\cite{Liu2004} (orange, amplitude scaled by a factor of 2.1). Reproduction from J. Appl. Phys. 95, 7637 (2004) with the permission of AIP Publishing (\href{https://doi.org/10.1063/1.1737046}{10.1063/1.1737046}). Reprinted with permission from L. A. Kappers, K. L. Sweeney, L. E. Halliburton, J. H. W. Liaw Phys. Rev. B 31, 6792 Copyright 1985 by the American Physical Society (\href{https://doi.org/10.1103/PhysRevB.31.6792}{10.1103/PhysRevB.31.6792})).\label{fig:figure1} }
\end{figure}

Figure~\ref{fig:figure1}b shows the laser-induced absorption change ${\alpha_\mathrm{li}=-\log(T_\mathrm{li}/T)/d}$ as a function of energy (blue data points) for the reduction annealed sample up to an exposure of $Q=I\cdot t\approx 2.4\cdot 10^{8} $Ws/m$^2$ (pump light polarization ${\bf E}\parallel {\bf c}$-axis).

Here, $T_\mathrm{li}$ and $T$ denote the transmission data with and without optical exposure.
We note that the data of figure~\ref{fig:figure1}b were determined without a polarizer in the optical paths of the spectrometer and are limited to the VIS/NIR spectral range of 0.5 eV$ - $2.75 eV (450 nm$ - $2\,500 nm) reported for the absorption of small polarons bound to Tantalum according to the data and results presented in Ref.~\cite{Kappers1985} (green data points) and Ref.~\cite{Liu2004} (orange data points). 

Obviously, a broad absorption band appears in the near-infrared region with a peak maximum at approximately 2.0 eV and a FWHM of 1 eV$-$1.2 eV. Compared to other small polaron absorption fingerprints, especially of small bound polarons in lithium niobate, the one in lithium tantalate features an unusually large width \cite{Schirmer2009}. Furthermore, on the low energy flank, a slightly pronounced shoulder is visible with a buckle around 1.5 eV. Both properties, width and shoulder, prevent the fitting of the absorption function according to the theory of small polarons. We note, that corresponding absorption bands were reported for c-LT by Kappers \textit{et al.} \cite{Kappers1985} and Liu \textit{et al.} \cite{Liu2004}, obtained with similar spectroscopic procedures. Overall, our measurements are in good agreement with literature data, especially along the low energy/long wavelength flank of the spectrum, where both, the broad width of the band and the shoulder at 1.5 eV can be retraced from literature. Disagreement between our measurement and the literature along the high energy flank of the spectrum ($\gg 2.2$ eV) is to be expected and likely caused by differences in the reduction treatment protocol.

\begin{figure}
\centering
\includegraphics[width=15cm]{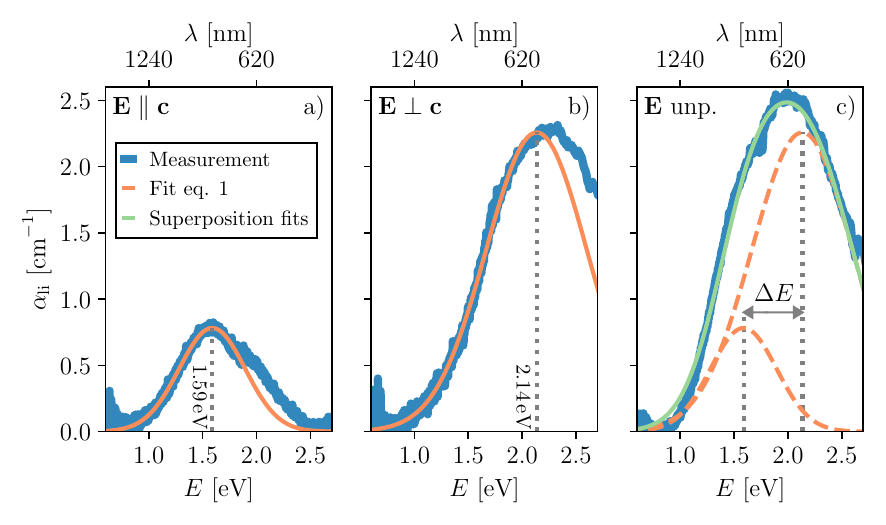}
\caption{Absorption spectra of c-LT in the illuminated state at $T=80$ K and using polarized probe light with a) ${\bf E}\parallel {\bf c}$-axis, and b) ${\bf E}\perp {\bf c}$-axis. The spectrum in c) is determined without polarizer in the optical paths. The orange lines represents fits of Equation~\ref{eq:equation1} to the data sets of a), b), that are used to reconstruct the spectrum in c) by minor amplitude scaling (green and orange dashed lines) with relative difference of $\approx 10$\%. 
\label{fig:figure2}}
\end{figure}

Performing the same measurements with linearly polarized probe light, ${\bf E}\parallel {\bf c}$- and ${\bf E}\perp {\bf c}$-axis, respectively, as shown in figure~\ref{fig:figure2}, reveals an anisotropic behaviour in the optical absorption of the gated polaronic states -- not reported in literature, so far.
In both cases, optical absorption bands are still identified. However, depending on the choice of the probe light polarization with respect to the polar $\bf{c}$-axis, spectral width and peak position differ from the unpolarized measurement to a large extent. In other words, the absorption feature originally measured with unpolarized light splits into two individual, clearly distinguishable absorption bands. For the case of $\bf{E}\perp\bf{c}$ a clear single peak is visible slightly above 2 eV, very close to the peak position of the unpolarized measurement of figure~\ref{fig:figure1}b. In direct comparison, however, the band is narrower in width and the apparent asymmetry on the low energy edge of the band is no longer visible. For the other polarization case, where $\bf{E}\parallel\bf{c}$, the peak position is shifted by about 0.5 eV and is located at 1.5 eV. Also here, no asymmetry can be seen regarding the low energy flank of the band. This peak is much weaker in absorption and shows only around 40 \% of the maximum absorption strength, relative to the band peaking at 2 eV. Considering their peak positions and relative strengths, the unpolarized case can therefore be considered as a superposed convolution of the polarization dependent bands.

Both absorption bands of figure~\ref{fig:figure2}a and figure~\ref{fig:figure2}b are fitted by the function 
\begin{equation}
\Delta\alpha\propto\frac{1}{\hbar\omega}\exp\left(-w(M-\hbar\omega)^2\right),
\label{eq:equation1}
\end{equation}
that was derived theoretically by D.~Emin for the description of small polaron optical absorption and that was already successfully used to deduce polaron formation energies from optical spectra of small polarons in lithium niobate \cite{Schirmer2009, Emin1975, Emin1993}. Here, $M$ is the peak position and $W=\sqrt{\log(2)/w}$ the HWHM of the undistorted gaussian band. We note that the model is fitted taking the data of the low energy flank of the band into account, i.e. $\ll 2.2$ eV. The parameters of the best fitting results shown in figure~\ref{fig:figure2} (orange lines) are summarized in  table~\ref{tab:table1}.

\begin{table}
\caption{Fit parameters of the polarized polaron bands according to equation~\ref{eq:equation1} with $M$ the peak position and $W$ the HWHM of the undistorted gaussian. $E_{\rm P}$ is the polaron binding energy and $\varepsilon$ the defect induced prelocalization energy. Assignment of the respective center in accordance with atomistic calculations (cf. figure~\ref{fig:figure4}) \label{tab:table1}}
\begin{tabular*}{\textwidth}{@{}c*{15}{@{\extracolsep{0pt plus
12pt}}c}}
\br
Species &   $M$ [eV]    &   $W$ [eV] &   $E_\mathrm{P}$ [eV] &   $\varepsilon$ [eV]  &   $W^2/M$ [eV]\\
\mr
Ta$_\mathrm{Li}^{4+}$                   &   2.23    &   0.54    &   1.05    &   0.14    &   0.13\\
Ta$_\mathrm{V}^{4+}$:V$_\mathrm{Li}$    &   1.65    &   0.35    &   0.45    &   0.74    &   0.08\\
\br
\end{tabular*}
\end{table}
In addition, the polaronic localization energies deduced in accordance with the molecular crystal model and its extensions \cite{Holstein1959, Holstein1959a, Appel1968, Austin1969, Schirmer2009} by $E_\mathrm{P}=1/8wE_\mathrm{vib}$ and $\varepsilon=M-2E_\mathrm{P}$ are given. 
For this purpose, the cryogenic temperature conditions justify to approximate the vibrational energy of the lattice to the zero point energy $E_\mathrm{vib}=\hbar\omega_\mathrm{ph}/2$ with the characteristic LO phonon energy $\hbar \omega_\mathrm{ph}\approx 0.1$ eV. Previous investigations into the optical fingerprints of polarons in polar oxides have shown that this is a reasonable choice \cite{Schirmer2009, Schirmer2006}. However, for the case of lithium tantalate this value is likely to represent an upper boundary of possible LO phonon energies \cite{Friedrich2016}. We note that the total electronic localization energies $E_\mathrm{P}+\varepsilon$ of the polarons are similar for both polaron species with values of $\approx 1.19$ eV, but slightly stronger localized than the Nb$_\mathrm{Li}^{4+}$ polaron in lithium niobate (0.95 eV) \cite{Guilbert2018}.

As demonstrated in figure~\ref{fig:figure2}c, it is possible to reconstruct the experimental data for the unpolarized measurement from the sum of both fits with very good agreement.

\subsection{Atomistic modeling}
\begin{figure}
\centering
\includegraphics[width=15 cm]{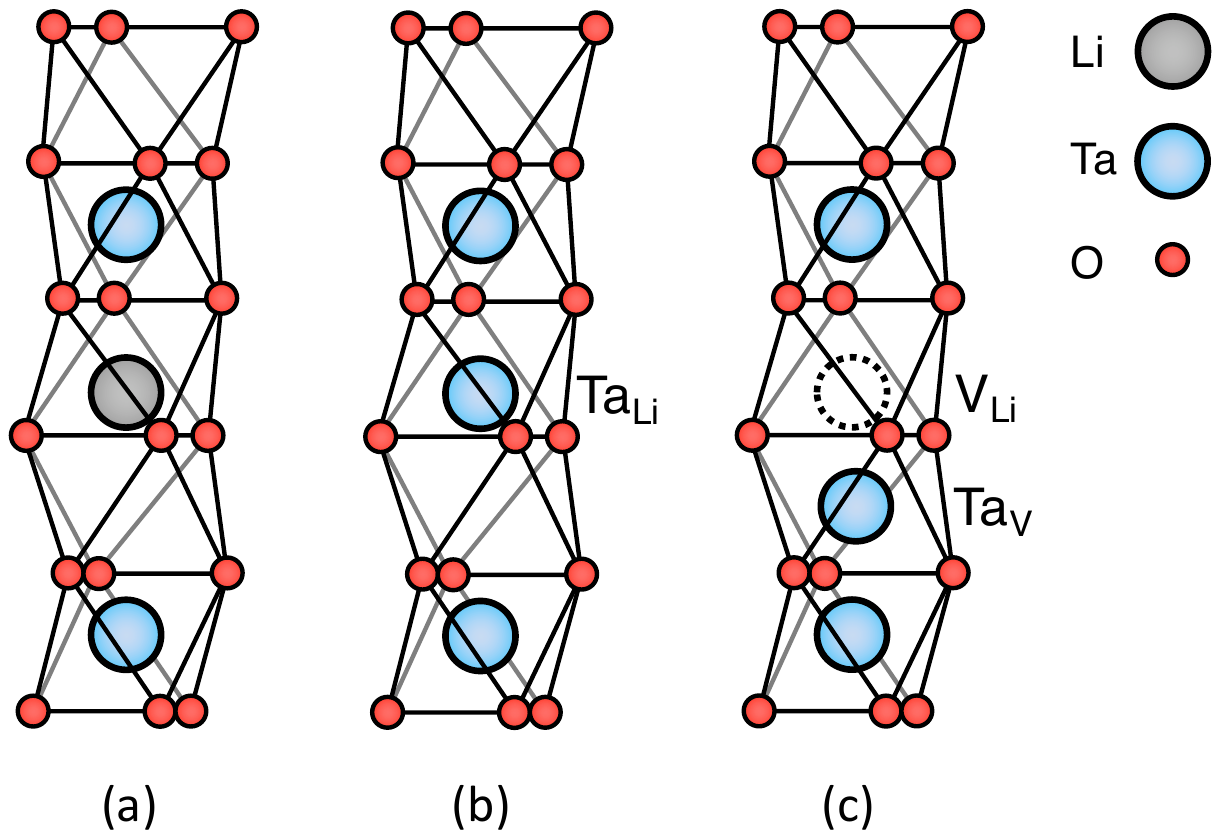}
\caption{\label{fig:figure3}Structural models of (a) defect free LiTaO$_3$, (b) Ta$_{\rm Li}^{5+}$ antisite, and (c) Ta$_{\rm V}^{5+}$:V$_{\rm Li}$ interstitial defect pair.
}
\end{figure}

 We have modeled small electron polarons in LiTaO$_3$ (i) bound to the Ta$_{\rm Li}^{5+}$ antisite defect (cf. figure~\ref{fig:figure3}b), and (ii) bound to the Ta$_{\rm V}^{5+}$ interstitial defect that is close to a lithium vacancy V$_{\rm Li}$ (cf. figure~\ref{fig:figure3}c), i.e. the Ta$_{\rm V}^{5+}$:V$_{\rm Li}$ defect pair, as structurally sketched in comparison to the defect free LiTaO$_3$ in figure~\ref{fig:figure3}a. We note that the interstitial defect can be also interpreted as a shifted antisite defect, so that both structures are stoichiometrically identic. Similar structures have been proposed in the past for LiNbO$_3$ \cite{Friedrich2017,Schmidt2021,Lerner1968,Zotov1994}, however, up to our knowledge they have not been used for atomistic calculations of small interstitial polarons in LiTaO$_3$ and/or for a respective comparison with small antisite polarons, so far. 

The atomic structures of the two investigated defect structures are optimized as described in section \ref{sec:methodology} and are modeled with and without the assumption of an excess electron localized at the defect site. As a result we find that the capture of an electron on a Ta$_{\rm Li}$ antisite (Ta$^{5+}_{\rm Li} \rightarrow $ Ta$^{4+}_{\rm Li}$) is accompanied by a large local lattice relaxation. In particular, the average Ta-O distance is increased by 2.8 \% from 2.022 \AA\, to 2.079 \AA. This distortion is a typical polaronic feature as it reflects the interaction of the localized charge with the surrounding lattice and is in very good agreement to the results presented in Ref.~\cite{Krampf2021}.

A rather similar behavior is found for the defect pair Ta$_{\rm V}^{5+}$:V$_{\rm Li}$, where the capture of an electron (Ta$^{5+}_{\rm V} \rightarrow $ Ta$^{4+}_{\rm V}$) leads to an expansion of the surrounding oxygen octahedra, as well. Here, we find an increase of the average Ta-O distance by 2.5 \% from 2.026 \AA\ to 2.077 \AA. Thus, the defect pair is compatible with the picture of a small bound Ta$_{\rm V}^{4+}$ polaron that induces a local lattice distortion. We remark that the structural relaxation of the neighboring V$_{\rm Li}$ is included in our atomistic models.

More important, we find that the formation energy of the structure for the small polaron bound to the interstitial defect is only 0.2 eV higher than the formation energy of the Ta$_{\rm Li}$ antisite and, therefore, is comparable with the one of the antisite polaron. It suggests that it is very likely that a non-vanishing amount of small Ta$_{\rm V}^{4+}$:V$_{\rm Li}$ polarons can be present in real samples at elevated temperatures. 

Straightforwardly, our calculations suggest that both models for bound polarons in c-LT allow for the formation of bipolarons upon a further capture of an electron. In particular, a hybrid is formed along the polar $\bf c$-axis between the Ta$_{\rm V}^{5+}$ defect level and an unoccupied orbital of the neighboring Ta$_{\rm Ta}^{5+}$ atom, whose energy is located within the conduction band. Such interstitial Ta$_{\rm V}^{4+}$:Ta$_{\rm Ta}^{4+}$ electron bipolaron is then very similar to the already established Ta$_{\rm Li}^{4+}$:Ta$_{\rm Ta}^{4+}$ electron bipolaron.

\begin{figure}
\centering
\includegraphics[width=15 cm]{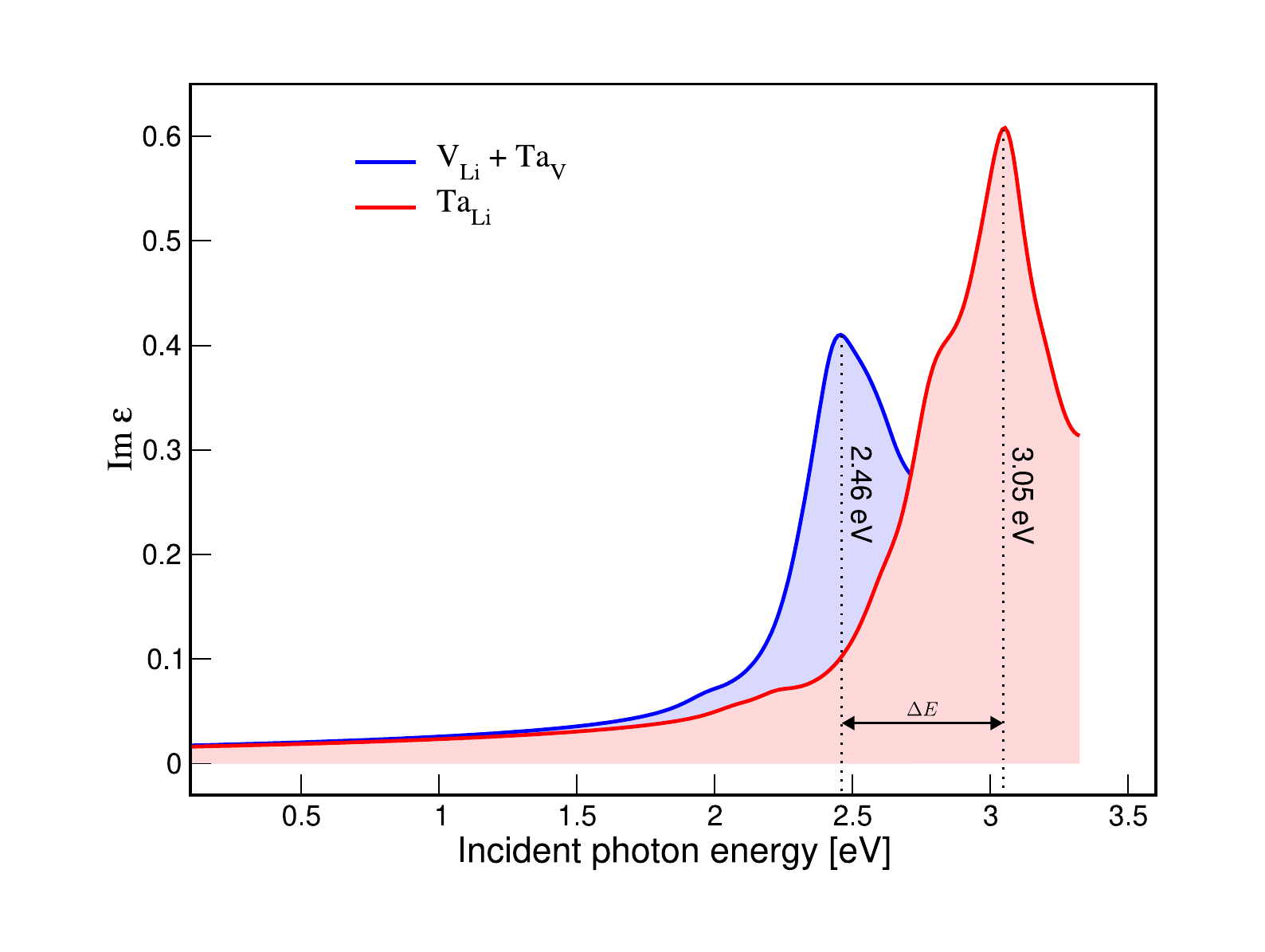}
\caption{\label{fig:figure4}Imaginary part of the dielectric function ($xx$-component) calculated at the BSE level of approximation for LiTaO$_{3}$ supercells modelling small bound polarons as a Ta$_{\rm Li}$ antisite (red line) or as a Ta$_{\rm V}$ interstitial close to a lithium vacancy V$_{\rm Li}$ (blue line). The polaronic signatures associated to the two different microscopic models are separated by about 0.6 eV.}
\end{figure} 

We now turn to the modeling results for the optical response of the described defect structures. In this case, and according to the experimental procedure for the illuminated state upon reduction annealing described above, the transitions are determined within the electronic structures of the two defects, but starting from the structural geometry of the respective bipolaron states. We here assume, that the bound polarons are formed upon optical dissociation of bipolarons (see also discussion below), and that the time scale at which the electronic transitions measured in the absorption spectra are recorded is orders of magnitude shorter than the lattice relaxation time. The linear optical response associated with the two defect models is quantified by the imaginary part of the dielectric function, that results in the modeled optical spectra shown in figure~\ref{fig:figure4}. For the sake of simplicity, we only plot the $xx$ tensor component. In the $zz$ direction the same optical features of both polarons are still visible, although with a slightly different distribution of the spectral weights. As a result, the two structural models reveal two distinct spectral features, that are centered at about 2.46 eV and 3.05 eV for the small bound Ta$_{\rm V}^{4+}$ (blue shaded area) and Ta$_{\rm Li}^{4+}$ (red shaded area) polarons, respectively, i.e., in the transparent region of the defect-free crystal.

\section{Discussion}
The experimental and theoretical consideration of interstitial Ta$_{\rm V}$:V$_{\rm Li}$ defect pairs for the self-localization of small electron polarons bound to defects unanimously lead us to the conclusion (i) that small bound Ta$_{\rm V}^{4+}$:V$_{\rm Li}$ electron polarons are probable to appear in LiTaO$_3$ single crystals grown from the congruent melt, and (ii) that small Ta$_{\rm V}^{4+}$:V$_{\rm Li}$ and small Ta$_{\rm Li}^{4+}$ electron polarons are both available after optical generation with distinct signatures in the optical spectra. 
\begin{figure}
\centering
\includegraphics[width=9.5cm]{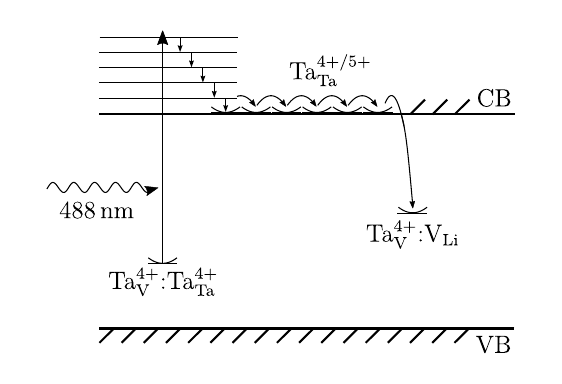}
\caption{\label{fig:figure5}Sketch of the optical gating mechanism in c-LT assuming the presence of Ta$_{\rm V}^{4+}$:Ta$_{\rm Ta}^{4+}$ bipolarons in the reduction annealed sample. For details see text.}
\end{figure} 
The respective separation of their optical fingerprints (cf.~figure~\ref{fig:figure2} and figure~\ref{fig:figure4}) is due to a difference of the polaron binding energies between interstitial and antisite polarons by about 0.5 eV, with a lower value for the interstitial case. Taking the widely accepted model for the optical generation of small bound Nb$_{\rm Li}^{4+}$ 
electron polarons in reduction annealed lithium niobate into account, we can straightforwardly suggest the following equivalent mechanism underlying the optical formation of small Ta$_{\rm V}^{4+}$:V$_{\rm Li}$ polarons in LiTaO$_3$ as schematically sketched in figure~\ref{fig:figure5}: The ground state of c-LT (as-grown sample of our study) already is characterized by the presence of both, Ta$_{\rm V}^{5+}$:V$_{\rm Li}$ interstitial and Ta$_{\rm Li}^{5+}$ antisite defects that are charge compensated by nearest neighboring lithium vacancies V$_{\rm Li}$. Furthermore, a considerable concentration of oxygen vacancies V$_{\rm O}$ is assumed. The reduction annealing process causes the incorporation of a large number of positively charged point defects into the crystal lattice \cite{Smyth1983, Mehta1991}. For reason of charge conservation the Fermi level is raised above the valence band edge. Electrons that occupy interband states are trapped at the variety of defect centers, most probably at the lowest lying Ta$_{\rm V}^{4+}$:Ta$_{\rm Ta}^{4+}$ and/or Ta$_{\rm Li}^{4+}$:Ta$_{\rm Ta}^{4+}$ bipolaron states. Bipolarons are characterized by optical absorption bands in the blue-green spectral range (e.g. centered at 2.5 eV in c-LN \cite{Schirmer2009}) and thus give rise to a greyish coloration of the single crystal. The change of the absorption spectra depicted in figure~\ref{fig:figure1} can be well attributed to the existence of bipolarons in reduction annealed c-LT. Using these samples, the photon energy of the incident laser light is able to excite an electron from the defect site to the conduction band, so that a small free Ta$_{\rm Ta}^{4+}$ polaron and a conduction band hot electron remain. This process is widely known as 'optical dissociation' or 'optical gating' of bipolarons \cite{Imlau2015, Schirmer2009}. The hot electron loses energy by its interaction with phonons until self-localization at the lowest edge of the Ta$_{\rm Ta}^{5+}$ valence band occurs. All these processes appear on the sub-ps time scale and end with the formation of small bound electron polarons within about 200 fs~\cite{Freytag2018,Krenz2022}. The relaxation back to its bipolaron state requires thermally activated hopping in the 3D lattice and is considerably decelerated by the low-temperature conditions of our experiments. Thus, the spectral investigation of the induced small polaron bands becomes possible and all individual experimental steps in our spectroscopic study, as well as the numerical results for the electronic structure and optical spectra can be fully understood with this approach. 

We note, however, that while the calculated separation of 0.5 eV between the polaronic signatures and their distance from the adsorption edge are in excellent agreement with the experiment, the absolute peak position in the calculations is blueshifted by about 1.0 eV with respect to the measurements.
In fact, the whole calculated spectrum (including the absorption edge) is blueshifted by about 1.0 eV with respect to the experiment. Indeed, the band gap energy is overestimated by 1.0 eV with respect to the experimentally determined value of E$_{\rm gap}=4.7$ eV. The latter is in good agreement with literature data for c-LT~\cite{He2008}. A discrepancy in the determination of the band gap energy between experiment and theory is not surprising and already was extensively discussed in one of our previous studies with LN as an example~\cite{Thierfelder2010}. It is due to an overestimation of the electronic gap in the GW calculation. Assuming that the data set of figure~\ref{fig:figure4} needs to be shifted by 1.0 eV to lower energies in order to correct the band gap energy, the absolute positions of the two appearing absorption bands are shifted to $\approx 1.5$ eV and $\approx2.0$ eV, i.e. in very good agreement with the experimental data and still differing by about 0.5 eV.

The difference in the ratio of peak position and spectral width (cf. $W^2/M$ values in table~\ref{tab:table1})
for Ta$_{\rm Li}^{4+}$ and Ta$_{\rm V}^{4+}$:V$_{\rm Li}$ points to a different influence of the defect potential on the polaron formation. Considering an upper limit for $W^2/M=0.14$ eV as predicted by Schirmer \textit{et al.}~\cite{Schirmer2009} (with $\hbar \omega_\mathrm{ph}\approx 0.1$ eV), the 
influence of the defect energy on the formation of the Ta$_\mathrm{Li}^{4+}$ polaron is minimal (compare also $W^2/M=0.09$ eV for the Nb$_\mathrm{Li}^{4+}$ antisite polaron in lithium niobate \cite{Schirmer2009}). In contrast,  the electron-lattice coupling for the interstitial Ta$_\mathrm{V}^{4+}$:V$_\mathrm{Li}$ polaron is strongly modified by the presence of the defect and the corresponding lattice deformation. We assume that this might be a result of the combined lattice distortion from the defect pair, i.e. of both the Ta$_\mathrm{V}$ and the neighboring V$_\mathrm{Li}$ defects. 

Taking our results and considerations into account, the question appears why the interstitial polaron state has not been considered in the description of the charge transport mechanisms of c-LT, before. Although we cannot make a definitive statement, this issue may have played only a subordinate role to date: The presence of interstitial defects has been considered without a doubt already within the original work of Kappers \textit{et al.}~\cite{Kappers1985}, but it focused on the role of oxygen vacancies in the reduction annealing process. The authors assigned a new magnetic resonance spectrum under optical gating conditions to the Ta ion valency change from $+5\rightarrow +4$, that accords with the assumption of either small bound Ta$_{\rm Li}^{4+}$ antisite and/or Ta$_{\rm V}^{4+}$:V$_{\rm Li}$ interstitial polarons. However, this result was used to exclude trapped electrons at the origin of the absorption change by reduction annealing, i.e. it was not applied to identify the defect origin of the induced absorption. As a result, the impact of using polarized probe light on the induced absorption peak at about 2.0 eV was not considered further. The same holds for the spectra depicted in the work of Liu \textit{et al.}~\cite{Liu2004}. Here, the authors focused on two-color photorefractivity in near-stoichiometric LT. Although an absorption peak at 2.2 eV appeared under optical gating conditions, no polarisation-dependent measurement was carried out. 

In contrast, in LN, the interstitial Nb$_{\rm V}$ defect with adjacent V$_{\rm Li}$~\cite{Zotov1994} was recently considered as metastable variant of the antisite Nb$_{\rm Li}$ defect in such a way that the antisite niobium potentially migrates to the next neighbouring unoccupied cation site~\cite{Schmidt2020}. This approach could be validated by theoretical calculations for at least some of the Nb$_{\rm Li}$ atoms, but also the formation of bipolarons localized at Nb$_{\rm V}$ seems likely~\cite{Schmidt2021}. The modelling results also coincide experimentally determined spectra of reduction annealed LN (cf e.g. Ref.~\cite{Schirmer2009}).

We finally like to discuss the structural origin of the polarization dependence of the light-induced small polaron absorption. It is reasonable to assume that the presence of a dichroitic small polaron absorption is a result of the local symmetry break of the lattice around the defect, which leads to the fact that the optical transition probabilities of the captured electron to one of the next-neighbouring ions differs for $\bf{E}\parallel c$ and $\bf{E}\perp c$. Such approach has been used to model the optical absorption of hole polarons in alkaline earth oxides with sixfold cubic symmetry at the example of MgO~\cite{Schirmer1974} and enables the successful reconstruction of the experimentally determined polarization dependent spectra. In particular, a splitting of the absorption feature with a energetic difference of about 0.6 eV was found. In our case of c-LT, we must additionally take into account the impact of light polarization of the pump light ${\bf E}\parallel \bf{c}$-axis used for optical gating of bipolarons. We can therefore not exclude the possibility that the polarized exposure up to a saturation state of the polaron population has influenced the symmetry breaking additionally and differently with respect to the antisite and interstitial defect.

\section{Conclusion}
In conclusion, the possibility to generate small electron polarons bound to interstitital defect sites in LiTaO$_3$ seems very likely, at least via optical dissociation of bipolarons. The formation energy is slightly lower in comparison with antisite polarons, so that its presence can be well assumed at room temperature, but with a considerably lower probability. The respective optical fingerprints overlap, so that a broad, slightly asymmetric absorption feature results, but also explains why the state has so far been largely ignored. More important is the impact of the additional small polaron state for the charge transport mechanism in LT, in particular for the 3D hopping transport process. The interstitial defect may act as additional intermediate electron trap, thus increasing the number of hopping events to a large extent as well as increasing the individual trapping duration. As a result, a strongly decelerated hopping transport mechanism must be expected that is of particular importance for the small-polaron mediated bulk photogalvanic effect. In more detail, the increase in the difference between conduction band and hopping transport can be used potentially to considerably increase the photogalvanic current density~\cite{Schirmer2011, Imlau2015}.

\section{Acknowledgment}
We gratefully acknowledge financial support by the Deutsche Forschungsgemeinschaft (DFG) through the research group FOR5044 (Grant No. 426703838, SA1948/3-1, IM37/12-1, SU12611/1-1, FR1301/42-1). Calculations for this research were conducted on the Lichtenberg high-performance computer of the TU Darmstadt and at the Höchstleistungsrechenzentrum Stuttgart (HLRS). The authors furthermore acknowledge the computational resources provided by the HPC Core Facility and the HRZ of the Justus-Liebig-Universität Gießen. 

\section*{References}
\bibliographystyle{iopart-num}
\bibliography{manuscript.bib}

\end{document}